\documentclass[fleqn,12pt]{wlscirep}
\usepackage[utf8]{inputenc}
\usepackage[T1]{fontenc}
\usepackage{amssymb,amsthm,amsmath,bbm,bbold,graphicx,epsfig,threeparttable,color}
\usepackage{ulem,enumerate} 
\usepackage{enumitem}
\usepackage[title]{appendix}
\usepackage{hyperref}

\newcommand{\ket}[1]{{|#1\rangle}}
\newcommand{\bra}[1]{{\langle#1|}}

\newcommand{\Tr}{{\rm Tr}}

\theoremstyle{definition}

\title{Precession-induced nonclassicality of the free induction decay of NV centers by a dynamical polarized nuclear spin bath}

\author[1]{Mu-Che Lin}
\author[2]{Ping-Yuan Lo}
\author[3,4]{Franco Nori}
\author[1,5,*]{Hong-Bin Chen}
\affil[1]{Department of Engineering Science, National Cheng Kung University, Tainan 701401, Taiwan}
\affil[2]{Department of Electrophysics, National Yang Ming Chiao Tung University, Hsinchu 300093, Taiwan}
\affil[3]{Quantum Computing Center, and Theoretical Quantum Physics Laboratory, Cluster for Pioneering Research, RIKEN, Wakoshi, Saitama 351-0198, Japan}
\affil[4]{Physics Department, University of Michigan, Ann Arbor, MI 48109-1040, USA}
\affil[5]{Center for Quantum Frontiers of Research \& Technology, NCKU, Tainan 701401, Taiwan}
\affil[*]{hongbinchen@gs.ncku.edu.tw}

\begin{abstract}
The ongoing exploration of the ambiguous boundary between the quantum and the classical worlds has spurred substantial developments in quantum science and technology. Recently, the nonclassicality of
dynamical processes has been proposed from a quantum-information-theoretic perspective, in terms of witnessing nonclassical correlations with Hamiltonian ensemble simulations. To acquire insights into
the quantum-dynamical mechanism of the process nonclassicality, here we propose to investigate the nonclassicality of the electron spin free-induction-decay process associated with an NV$^-$
center. By controlling the nuclear spin precession dynamics via an external magnetic field and nuclear spin polarization, it is possible to manipulate the dynamical behavior of the electron spin,
showing a transition between classicality and nonclassicality. We propose an explanation of the classicality-nonclassicality transition in terms of the nuclear spin precession axis orientation and
dynamics. We have also performed a series of numerical simulations supporting our findings. Consequently, we can attribute the nonclassical trait of the electron spin dynamics to the behavior of
nuclear spin precession dynamics.
\end{abstract}
\begin{document}

\flushbottom
\maketitle


\section{Introduction}

Along with the development of quantum theory, the ongoing exploration of the ambiguous boundary between the quantum and the classical worlds has attracted extensive interest
\cite{ballentine_q_statistics_rmp_1970,zurek_q-c_transition_rmp_2003,schlosshauer_q-c_transition_rmp_2005,modi_q-c_boundary_discord_rmp_2012}.
Although the intuitive viewpoints, e.g., local realism or commutativity of conjugated observables, are seemingly natural and valid in the classical world, they may result in contradicting predictions in
the quantum realm. Therefore, the failure of classical strategies attempting to explain an experimental outcome can be conceived as a convincing evidence of quantum nature beyond classical intuition, or
nonclassicality.

One of the most famous paradigms is the experimental violation \cite{aspect_bell_test_prl_1981,hensen_bell_test_nature_2015,giustina_bell_test_prl_2015,shalm_bell_test_prl_2015} of Bell's inequality
\cite{bell_ineq_phys_1964}, which is derived under the assumptions of realism and locality. With the explicit violation of Bell's inequality, the bipartite correlation demonstrates a genuinely
nonclassical nature, i.e., Bell nonlocality \cite{brunner_bell_nonlocal_rmp_2014}, which can never be explained classically in terms of the local hidden variable model. Following the same logic, the
nonclassicality of a bosonic field is characterized by the Wigner function \cite{wigner_func_pr_1932} or the Glauber-Sudarshan $P$ representation \cite{glauber_pr_1963, sudarshan_prl_1963,
PhysRevA.82.013824,PhysRevA.83.053814,PhysRevA.91.042309,PhysRevA.92.062314}. The negativity in these functions demonstrates nonclassicality with a phase space description.

Additionally, there is another emerging type of nonclassicality considering the nature of quantum dynamical processes. Considerable efforts have been devoted to approaching this problem
\cite{neill_n_cla_trans_prl_2010,rahimikeshari_process_n_cla_prl_2013,krishna_process_n_cla_pra_2016,jenhsiang_process_n_cla_scirep_2017,george_coh_wit_pra_2018,smirne_process_n_cla_qst_2019,
alireza_process_n_cla_prl_2022}. Noteworthily, these approaches focus on certain nonclassical properties of interest of the quantum systems and continuously monitor their temporal evolutions as
indicators of dynamical process nonclassicality. Recently, it has been proposed \cite{hongbin_process_n_cla_prl_2018,hongbin_process_n_cla_nc_2019,hongbin_cher_sr_2021} an alternative definition of
dynamical process nonclassicality, based on the failure of the classical strategy to simulate the incoherent dynamical process. The classical strategy is formulated in terms of a Hamiltonian ensemble (HE)
\cite{kropf2016effective,hongbin_disordered_sr_2022}, which was initially devoted to investigating the decoherence induced by a disordered medium or classical noise \cite{gneiting2016incoherent,
kropf2017effective,gneiting_disordered_prl_2017,kropf_disordered_prr_2020}. The classicality behind the HE would become clear after looking for additional insights into the cause of the incoherent
dynamics.

Quantum systems inevitably interact with their surrounding environments \cite{breuer_textbook,weiss_textbook,breuer_non_mark_review_rmp_2016,ines_non_mark_review_rmp_2017}; meanwhile, complicated
correlations will be established between the systems and their environments during these interactions. These correlations are typically fragile and transient, as the quantum systems are subject to the
fluctuations caused by a huge number of environmental degrees of freedom. As a result, the vanishing of the correlations constitutes one of the main sources of decoherence, leading to incoherent
dynamical processes. Therefore, a natural way to classify decoherence is based on the properties of the correlations established during the interactions. However, such a naive classification is not
feasible due to the huge number of environmental degrees of freedom, rendering the environments, as well as the established correlations, inaccessible.

On the other hand, it has been shown that the classically correlated system-environment correlation results in an incoherent dynamics admitting HE simulations \cite{hongbin_process_n_cla_prl_2018}. This
implies that the failure of HE simulations serves as a witness for the establishment of nonclassical correlations during the interactions. Such incoherent dynamics violating HE simulations necessarily
appeals to nonclassical correlations, rather than reproduced classically by statistical noise. We therefore proposed to classify an incoherent dynamical process according to the possibility to
simulate the reduced system dynamics with HEs. It should be stressed that, in this definition, the actual system-environment correlations are deliberately ignored as they are inaccessible to the
reduced system exclusively; meanwhile, we attempt to reproduce the incoherent effects on the reduced system classically by using HE simulations.

Additionally, this HE-simulation approach can be further promoted to a representation of the incoherent dynamics over the frequency domain, referred to as the canonical Hamiltonian ensemble
representation (CHER) \cite{hongbin_process_n_cla_nc_2019}. This is reminiscent of the conventional Fourier transform, transforming a temporal sequence into its frequency spectrum. In contrast, CHER is
defined in accordance with the underlying algebraic structure of the HE, highlighting its difference from the conventional one by the non-Abelian algebraic structure \cite{hongbin_disordered_sr_2022}.
Along with the quasi-distribution of the CHER, we can define a quantitative measure of nonclassicality \cite{hongbin_process_n_cla_nc_2019}. Moreover, we place particular emphasis on the practical
viability, as this approach relies only on the information of the reduced system, irrespective of the inaccessible environment.

Although dynamical process nonclassicality has been investigated from a quantum-information-theoretic perspective, its quantum-dynamical mechanism is still not understood. To acquire insights into the
system-environment interactions giving rise to the classicality-nonclassicality transition, meanwhile underpinning the practical viability of this appraoch, we will discuss the CHER of the
free-induction-decay (FID) process of the electron spin associated with a single negatively charged nitrogen-vacancy (NV$^-$) center in diamond. Due to its unique properties, particularly its long coherence time even at room temperature \cite{kennedy2003long,herbschleb_nv_t2_nc_2019,balasubramanian_nv_iso_eng_nat_mater_2009,ishikawa_nv_iso_eng_nl_2012,maurer_nv_center_science_2012}, NV$^-$
centers are a promising candidate for applications in various branches of quantum technologies, ranging from quantum information processing \cite{maurer_nv_center_science_2012,dutt2007quantum,
zagoskin_nv_center_prb_2007,ladd_nv_q_comp_nature_2010,buluta_rep_pro_phys_2011,nakazato_nv_center_comm_phys_2022}, highly-sensitive nanoscale magnetometry \cite{schmitt_nv_magnetometry_science_2017,
casola2018} and electrometry \cite{dolde_nat_phys_2011,dolde2014nanoscale}, bio-sensing in living cells \cite{mcguinness_nv_bio_sen_nat_nano_2011,petrini2020quantum}, emerging quantum materials
\cite{li_q_device_nv_center_prl_2016,li_q_device_nv_center_prapp_2018,ai_q_device_nv_center_prb_2021}, to test bed of fundamental quantum physics \cite{lu_info_back_flow_nv_prl_2020}. The primary source
of decoherence of the electron spin comes from the hyperfine interaction to the nuclear spin bath of carbon isotopes $^{13}$C randomly distributed in the diamond lattice. Therefore, several techniques
have been developed for prolonging the coherence time by engineering the nuclear spin bath, including the isotopic purification
\cite{balasubramanian_nv_iso_eng_nat_mater_2009,ishikawa_nv_iso_eng_nl_2012,maurer_nv_center_science_2012} and dynamic nuclear spin polarization (DNP)
\cite{takahashi_nuc_spin_pola_prl_2008,london_nuc_spin_pola_prl_2013}.

Here we investigate how the nonclassicality of the FID process is induced by the nuclear spin precession dynamics. Since the external magnetic field and the nuclear spin polarization are two
experimentally controllable mechanisms manipulating the nuclear spin precession dynamics, we found that the dynamical behavior of FID and the corresponding CHER are sensitive to both controllable
properties. We have also observed a transition between classicality and nonclassicality by engineering the nuclear spin bath via the polarization orientation, particularly the $x$ component of the
polarization. Consequently, we can attribute the nonclassical traits of the electron spin FID process to the behavior of nuclear spin precession dynamics based on the response to two controllable
properties. Finally, in order to underpin the experimental viability of our numerical simulation, we also present an experimental pulse sequence for carrying out the model.

\section{Theory of dynamical process nonclassicality}

\subsection{Decoherence under Hamiltonian ensembles}

\begin{figure}[ht]
\centering
\includegraphics[width=\textwidth]{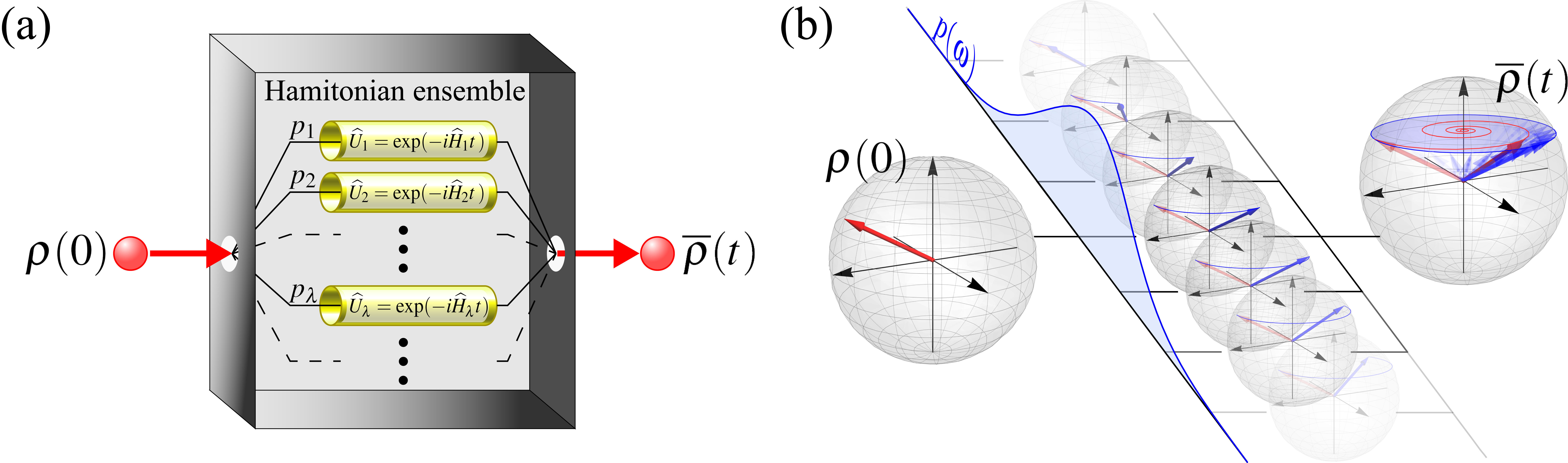}
\caption{Schematic illustration of the ensemble-averaged dynamics under HEs. (a) A HE $\{(p_{\lambda},\widehat{H}_{\lambda})\}_{\lambda}$ consists of a collection of traceless Hermitian
operators $\widehat{H}_{\lambda}\in\mathfrak{su}(n)$ associated with a probability $p_{\lambda}$ of occurrence. An initial state $\rho(0)$ will be randomly assigned to a certain unitary channel
$\widehat{U}_{\lambda}(t) = \exp(-i\widehat{H}_{\lambda}t)$ according to $p_\lambda$, giving rise to an ensemble-averaged dynamics $\overline{\rho}(t)$ described by
Eq.~(\ref{eq_ensemble-averaged-dynamics}). (b) For the HE of spectral disorder $\{(p(\omega),\omega\hat{\sigma}_z/2)\}_\omega$, every unitary operator $\exp(-i\omega\hat{\sigma}_z t/2)$ rotates
the qubit state about the $z$-axis of the Bloch sphere at a fluctuating angular frequency $\omega$. Hence the qubit is subject to an uncertainty described by the probability distribution $p(\omega)$,
and the ensemble-averaged dynamics $\overline{\rho}(t)$ is a pure dephasing described by Eq.~(\ref{eq_pure_dephasing}). The incoherent dynamical behavior is determined by $p(\omega)$ via the Fourier
transform $\phi(t)=\int p(\omega)\exp(-i\omega t)d\omega$.}
\label{fig_illustration}
\end{figure}

The central role in our approach, modeling an incoherent dynamics, is played by the Hamiltonian ensemble (HE) $\{(p_{\lambda},\widehat{H}_{\lambda})\}_{\lambda}$, which consists of a collection of
traceless Hermitian operators $\widehat{H}_{\lambda}\in\mathfrak{su}(n)$ associated with a probability $p_{\lambda}$ of occurrence [figure~\ref{fig_illustration}(a)]. The index $\lambda$ could be very general and may be continuous and/or a multi-index. Every member Hamiltonian operator $\widehat{H}_{\lambda}$ acts on the same system Hilbert space, leading to a unitary time-evolution operator
$\widehat{U}_{\lambda}(t) = \exp(-i\widehat{H}_{\lambda}t)$. Then a HE will randomly assign an initial state $\rho(0)$ to a certain unitary channel $\widehat{U}_{\lambda}(t)$ according to $p_\lambda$,
giving rise to an ensemble-averaged dynamics described by
\begin{equation}
\overline{\rho}(t)=\mathcal{E}_{t}\{\rho(0)\}=\int p_{\lambda}\widehat{U}_{\lambda}(t)\rho(0)\widehat{U}^{\dag}_{\lambda}(t)d\lambda.
\label{eq_ensemble-averaged-dynamics}
\end{equation}
Irrespective of the unitarity of each single channel, the ensemble-averaged dynamics (\ref{eq_ensemble-averaged-dynamics}) demonstrates an incoherent behavior due to the averaging procedure over all
unitary realizations \cite{kropf2016effective,hongbin_disordered_sr_2022,gneiting2016incoherent}. For instance, when a single qubit is subject to spectral disorder with a HE given by
$\{(p(\omega),\omega\hat{\sigma}_z/2)\}_\omega$, where $p(\omega)$ can be any probability distribution function [figure~\ref{fig_illustration}(b)], it undergoes a pure dephasing dynamics given by
\begin{equation}
\overline{\rho}(t)=\int_{-\infty}^{\infty}p(\omega)e^{-i\omega\hat{\sigma}_{z}t/2}\rho_{0}\,e^{i\omega\hat{\sigma}_{z}t/2}d\omega=\begin{bmatrix}
\rho_{\upuparrows} & \rho_{\uparrow\downarrow}\,\phi(t) \\
\rho_{\downarrow\uparrow}\,\phi^{\ast}(t) & \rho_{\downdownarrows}
\end{bmatrix}
\label{eq_pure_dephasing}
\end{equation}
with the dephasing factor $\phi(t)=\int p(\omega)\exp(-i\omega t)d\omega$ being the Fourier transform of $p(\omega)$.

Interestingly, the heuristic example shown in Eq.~(\ref{eq_pure_dephasing}) provides further insights into the role played by HEs. On the one hand, the probability distribution $p(\omega)$ completely
determines the pure dephasing via the Fourier transform in Eq~(\ref{eq_pure_dephasing}). Similar situations occur when one considers the case of multivariate probability distribution
\cite{hongbin_process_n_cla_nc_2019,hongbin_disordered_sr_2022}. Therefore, the probability distribution $p_\lambda$ encapsulated within a HE can be conceived as a characteristic representation of the
incoherent dynamics, which is even faithful for the case of pure dephasing of any dimension \cite{hongbin_process_n_cla_nc_2019,hongbin_cher_sr_2021}. This observation endows the probability
distribution $p_\lambda$ a new interpretation as a representation function of incoherent dynamics over the frequency domain, referred to as canonical Hamiltonian ensemble representation (CHER).

On the other hand, since every unitary operator $\exp(-i\omega\hat{\sigma}_z t/2)$ in Eq.~(\ref{eq_pure_dephasing}) can be interpreted geometrically as a rotation about the $z$-axis of the Bloch sphere
at angular frequency $\omega$, the pure dephasing is a result of the accumulation of a random phase, in line with the conventional interpretation of pure dephasing in the manner of the random phase
model [figure~\ref{fig_illustration}(b)]. Consequently, the ensemble-averaged dynamics~(\ref{eq_ensemble-averaged-dynamics}) under a HE can be considered as a statistical mixture of various unitary
rotations weighted by $p_\lambda$. This underpins the classicality behind the HE.

\subsection{Canonical Hamiltonian ensemble representation}

Before further exploring the characterization of dynamical process nonclassicality with CHER, we elucidate how can a HE be recast into a Fourier transform using the formalism of group theory. This also
strengthens the formal connection of the CHER to an incoherent dynamics beyond the above empirical observation.

As every member Hamiltonian operator $\widehat{H}_\lambda\in\mathfrak{su}(n)$ is Hermitian, it can be expressed as a linear combination
\begin{equation}
\widehat{H}_\lambda=\sum_{m=1}^{n^2-1}\lambda_m\widehat{L}_m=\vec{\lambda}\cdot\widehat{\boldsymbol{L}}
\label{eq_operator_expansion}
\end{equation}
of $n^2-1$ traceless Hermitian generators $\widehat{L}_m$ of $\mathfrak{su}(n)$. Therefore, the index $\lambda$ parameterizing the HE is an $(n^2-1)$-dimensional real vector $\vec{\lambda}$.
Here we restrict ourselves to traceless member Hamiltonian operators $\widehat{H}_\lambda$ without loss of generality, because the trace plays no role in the ensemble-averaged
dynamics~(\ref{eq_ensemble-averaged-dynamics}) due to the commutativity $[\widehat{I},\widehat{L}_m]=0$ of the identity operator $\widehat{I}$ to all $\widehat{L}_m\in\mathfrak{su}(n)$.

The commutator is critical for a Lie algebra as it determines many properties and the algebraic structure to a very large extent. For example, the generators of $\mathfrak{su}(n)$ should satisfy
$[\widehat{L}_k,\widehat{L}_l]=i2c_{klm}\widehat{L}_m$, where the $c_{klm}$'s are called structure constants, satisfying the relation $c_{klm}=-c_{lkm}=-c_{mlk}$.
Additionally, the commutator can be used to induce the adjoint representation of $\mathfrak{su}(n)$ of fundamental importance according to
\begin{equation}
ad:\widehat{L}_m\mapsto\widetilde{L}_m=[\widehat{L}_m,\quad].
\end{equation}
Namely, the adjoint representation $ad$ uniquely associates each generator $\widehat{L}_m$ to an endomorphism $\widetilde{L}_m:\mathfrak{u}(n)\rightarrow\mathfrak{u}(n)$ with its action defined by the
commutator as $\widetilde{L}_m\{\widetilde{L}_k\}:=[\widehat{L}_m,\widetilde{L}_k]$. Since each endomorphism $\widetilde{L}_m$ itself is a linear map, it admits a matrix representation of
$n^2\times n^2$ dimension with elements given by the structure constants and its action is described by the ordinary matrix multiplication. Furthermore, according to the linear
combination~(\ref{eq_operator_expansion}) and the bilinearity of the commutator, each member Hamiltonian can be associated according to
\begin{equation}
ad:\widehat{H}_\lambda\mapsto\widetilde{H}_\lambda=[\widehat{H}_\lambda,\quad]=\sum_{m=1}^{n^2-1}\lambda_m\widetilde{L}_m=\vec{\lambda}\cdot\widetilde{L}.
\end{equation}
Along with the adjoint representation $ad$, one can show that the action of each single unitary channel in Eq.~(\ref{eq_ensemble-averaged-dynamics}) can be recast into an exponential form with
respect to the generators $\widetilde{L}_m$ as
\begin{equation}
\exp(-i\widehat{H}_\lambda t)\rho\exp(i\widehat{H}_\lambda t)=\sum_{j=0}^{\infty}\frac{(-it)^j}{j!}[\widehat{H}_\lambda,\rho]_{(j)}=\exp(-i\vec{\lambda}\cdot\widetilde{L}t)\{\rho\},
\end{equation}
where the multiple-layer commutator is defined iteratively according to $[\widehat{H}_\lambda,\rho]_{(j)}=[\widehat{H}_\lambda,[\widehat{H}_\lambda,\rho]_{(j-1)}]$ and
$[\widehat{H}_\lambda,\rho]_{(0)}=\rho$.

From the above equations, for a given HE $\{(p_\lambda,\widehat{H}_\lambda)\}_{\lambda}$, we can recast  \cite{hongbin_process_n_cla_nc_2019} the right hand side of
Eq.~(\ref{eq_ensemble-averaged-dynamics}) into a Fourier transform expression from the probability distribution $p_\lambda$, on a locally compact group $\mathcal{G}$ parameterized by
$\lambda=\vec{\lambda}\in\mathbb{R}^{n^2-1}$, to the dynamical linear map $\mathcal{E}_t^{(\widehat{L})}$:
\begin{equation}
\mathcal{E}_t^{(\widetilde{L})}=\int_\mathcal{G} p_\lambda\exp(-i\lambda\widetilde{L}t)d\lambda. \label{eq_ft_on_group}
\end{equation}
Meanwhile, the action of the incoherent dynamics $\mathcal{E}_t$ on a density matrix $\rho$ can be expressed in terms of ordinary matrix multiplication
\begin{equation}
\mathcal{E}_t\{\rho\}\Rightarrow\mathcal{E}_t^{(\widehat{L})}\cdot\rho,
\label{eq_action_of_dynamical_map}
\end{equation}
with $\rho$ on the left hand side being an $n\times n$ density matrix, while the one on the right hand side being an $n^2$-dimensional real vector
$\rho=\{n^{-1},\vec{\rho}\}$ in the sense of the linear combination $\rho=n^{-1}\widehat{I}+\vec{\rho}\cdot\widehat{\boldsymbol{L}}$.

In summary, Eq.~(\ref{eq_ft_on_group}) associates a probability distribution $p_\lambda$ within a HE to the incoherent dynamics $\mathcal{E}_t$ under HE, i.e.,
\begin{equation}
p_\lambda\mapsto\mathcal{E}_t^{(\widetilde{L})},
\end{equation}
via the Fourier transform using the formalism of group theory. This explicitly elucidates that the role of $p_\lambda$ as a characteristic representation over the frequency domain for $\mathcal{E}_t$,
referred to as CHER. Additionally, the particular versatility of CHER lies in characterizing and quantifying the dynamical process nonclassicality. To do this, we will replace the probability distribution
$p_\lambda$ with the quasi-distribution $\wp_\lambda$, to incorporate the possibility that $\wp_\lambda$ may contain negative values, which serve as an indicator of the nonclassical nature in the
dynamical process $\mathcal{E}_t$. The emergence of negative values will be clarified in the following discussion.

\subsection{Dynamical process nonclassicality}

Upon elucidating how a given HE induces an incoherent dynamics, we consider a reverse problem of simulating a given incoherent dynamics with HEs, which servers as the classical strategy in
the characterization of dynamical process nonclassicality.

As discussed in the Introduction, the incoherent behavior of an open system dynamics is caused by the destruction of the correlations established during the system-environment interaction. However,
these correlations are typically not fully accessible in an experiment; therefore, it is not feasible to precisely infer whether an incoherent behavior results from quantum or classical correlations.
Whereas, the reduced system dynamics is, in principle, fully attainable with the technique of, e.g., quantum process tomography (QPT) experiments or theoretically solving a master equation
\cite{silbey_comparison_cjp_2001,hongbin_3sbm_scirep_2015,chruscinski_open_system_dyna_osid_2017}. Consequently, in our approach, we focus exclusively on the reduced system dynamics and ignore
the obscure actual system-environment correlations; meanwhile we attempt to explain the dynamics classically with HE simulations.

The classicality behind the HE simulations can be understood by recalling that \cite{hongbin_process_n_cla_prl_2018}, if the system and its environment can at most establish classical correlations,
without quantum discord nor entanglement, during their interactions, then the reduced system dynamics corresponds to a HE. This means that, if a given incoherent dynamics admits a HE simulation, then
one cannot tell it apart from a classical model reproducing exactly the same dynamics relying merely on classical correlations. In other words, the given incoherent behavior can be explained classically
by a statistical mixture of a collection of unitary channels.

On the contrary, if one fails to construct the simulating HEs, then the incoherent dynamics should go beyond classical HEs, showcasing the
nonclassicality of the dynamical process. The nonexistence of simulating HEs can be proven by the necessity to resort to a nonclassical HE accompanied by a negative quasi-distribution $\wp_\lambda$.
This renders the CHER $\wp_\lambda$ quite versatile, not only representing the incoherent dynamics but also characterizing the dynamical process nonclassicality.

The nonclassicality is defined from a quantum-information-theoretic perspective, i.e., the inference of nonclassical system-environment correlations. However, a quantum-dynamical viewpoint
providing insights into the origin of nonclassicality is still obscure. In the following, we will discuss how to implement our approach with an authentic quantum system. This in turn reveals
the mechanism of system-environment interactions giving rise to nonclassical dynamics and the classicality-nonclassicality transition caused by the environmental dynamics.

\section{Dynamics of NV$^-$ centers}

\subsection{Theoretical model}

\begin{figure}[ht]
\centering
\includegraphics[width=\textwidth]{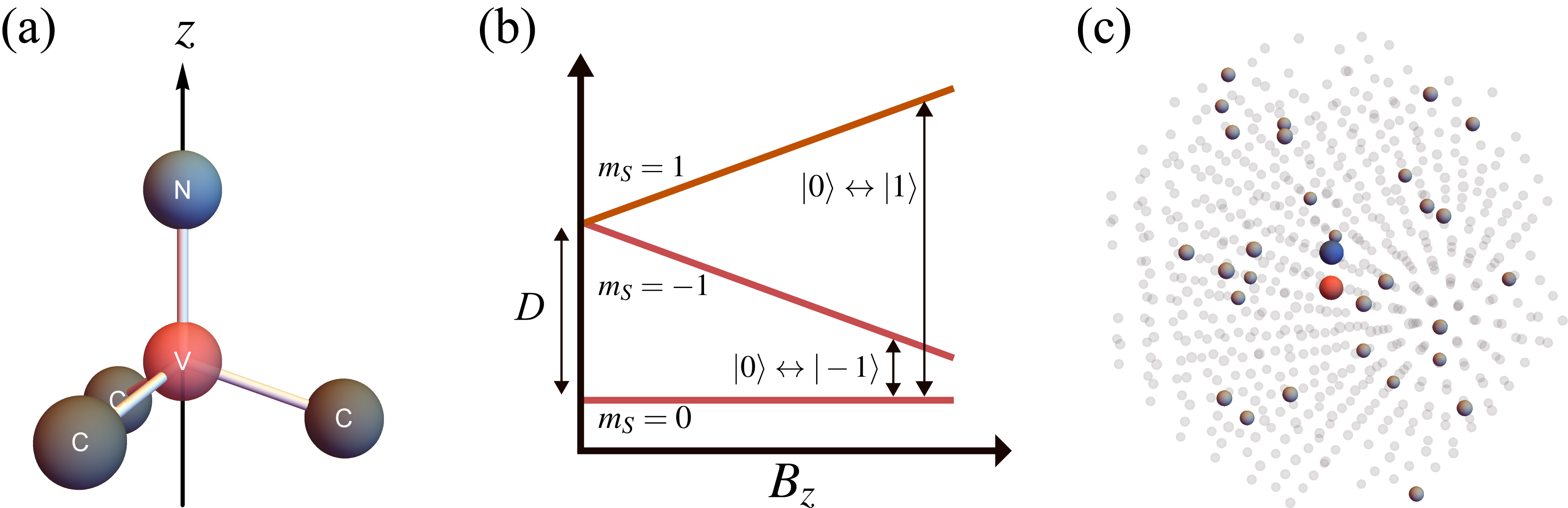}
\caption{Atomic structure and ground state energy level of an NV$^-$ center in a diamond lattice.
(a) The atomic structure of an NV$^-$ center in a diamond lattice, which consists of a substitutional nitrogen (N) and a vacancy (V) in an adjacent lattice site. The $C_3$ rotation axis defines
an intrinsic $z$-axis for the electron spin. (b) The ground state of the electron spin is a spin-1 triplet with a zero-field splitting $D$ between sublevels $m_S=0$ and $m_S=\pm 1$. An external
magnetic field will further lift the degeneracy between sublevels $m_S=\pm 1$. Then one can selectively address the two spin transitions $\ket{0}\leftrightarrow\ket{\pm 1}$ with an appropriate MW
pulse, forming a logical qubit. (c) The most abundant species in the diamond lattice is the spinless $^{12}$C nucleus (light gray sphere), which do not interact with the electron spin. The
primary source of decoherence comes from the randomly distributed $^{13}$C isotopes (dark gray spheres) with nuclear spin $J=1/2$. Due to the low concentration of $^{13}$C (1.1$\%$ natural abundance),
the nuclear dipole-dipole interaction and the electron-nucleus Fermi contact are negligible.}
\label{fig_nv_introduction}
\end{figure}

A single negatively charged nitrogen-vacancy (NV$^-$) center is a point defect in diamond consisting of a substitutional nitrogen (N) and a vacancy (V) in an adjacent lattice site
[figure~\ref{fig_nv_introduction}(a)]. The $C_3$ rotation axis defines an intrinsic $z$-axis for the electron spin. The NV$^-$ center has an electron spin $S=1$ triplet as its ground state with a
zero-field splitting $D/2\pi=2.87$ GHz between sublevels $m_S=0$ and $m_S=\pm 1$. By applying an external magnetic field $\vec{B}$, the degeneracy between $m_S=\pm 1$ can be lifted due to the Zeeman
splitting [figure~\ref{fig_nv_introduction}(b)]. Then the two different single spin transitions $\ket{0}\leftrightarrow\ket{\pm 1}$ can be addressed by selective microwave (MW) excitations. The
lattice sites are mostly occupied by the spinless $^{12}$C nuclei [light gray spheres in figure~\ref{fig_nv_introduction}(c)], while the electron spin decoherence is caused by the randomly distributed
$^{13}$C isotopes [dark gray spheres in figure~\ref{fig_nv_introduction}(c)] with nuclear spin $J=1/2$. The natural abundance of $^{13}$C is about 1.1$\%$, leading to a spin qubit relaxation time $T_1$
in the order of milliseconds \cite{redman_nv_relaxation_prl_1991,neumann_nv_entanglement_science_2008} and a dephasing time $T_2^\ast$ of microseconds \cite{childress_nv_center_science_2006,
liu_nv_center_fid_sr_2012,maze_nv_center_fid_njp_2012}. The $^{13}$C concentration can even be depleted in isotopically purified samples to prolong the coherence time
\cite{balasubramanian_nv_iso_eng_nat_mater_2009,ishikawa_nv_iso_eng_nl_2012,maurer_nv_center_science_2012}.

In the presence of an external magnetic field $\vec{B}$, the total Hamiltonian $\widehat{H}_\mathrm{T}=\widehat{H}_\mathrm{NV}+\widehat{H}_\mathrm{C}+\widehat{H}_\mathrm{I}$ consists of three terms.
The first term describes the free Hamiltonian of the electron spin associated to the NV$^-$ center,
\begin{equation}
\widehat{H}_\mathrm{NV}=D\widehat{S}^2_z+\gamma_e\vec{B}\cdot\widehat{S},
\end{equation}
with $D/2\pi=2.87$ GHz being the zero-field splitting and $\gamma_e/2\pi=2.8025$ MHz/G the electron gyromagnetic ratio. The second term describes the nuclear spin bath consisting of $^{13}$C isotopes of
$J=1/2$ indexed by $k$:
\begin{equation}
\widehat{H}_\mathrm{C}=\sum_k\gamma_\mathrm{C}\vec{B}\cdot\widehat{J}^{(k)},
\end{equation}
with $\gamma_\mathrm{C}/2\pi=1.0704$ kHz/G being the gyromagnetic ratio of the $^{13}$C nuclei. Note that we have neglected the dipole-dipole interaction between $^{13}$C nuclei as its effect is much
slower than the electron spin decoherence considered here. The last term $\widehat{H}_\mathrm{I}$ takes into account the hyperfine coupling between electron spin and nuclear spin, which in general
includes two contributions; namely, the Fermi contact and the dipole-dipole interaction. The former is proportional to the overlap of the electron wavefunction at the position of a nucleus. Since the
electron wavefunction is highly localized at the defect, this effect is negligible for nuclei farther away than 5 {\AA} from the NV$^-$ center. In our simulation, we have confirmed the relevance of the
dipole-dipole hyperfine interaction as well as the negligibility of the Fermi contact by post-selecting a randomly generated configuration with all $^{13}$C nuclei lying outside this radius of 5 {\AA}.
Therefore, we consider the dipole-dipole interaction to the $k^\mathrm{th}$ nucleus exclusively with
\begin{equation}
\widehat{H}_\mathrm{I}=\widehat{S}\cdot\sum_k\stackrel{\leftrightarrow}{A}^{(k)}\cdot\widehat{J}^{(k)},
\label{eq_int_hamiltonian}
\end{equation}
where the hyperfine coefficients are given by
\begin{equation}
A_{ij}^{(k)}=\alpha^{(k)}\left[\vec{e}_i\cdot\vec{e}_j-3(\vec{e}^{(k)}\cdot\vec{e}_i)(\vec{e}^{(k)}\cdot\vec{e}_j)\right]
\end{equation}
with
\begin{equation}
\alpha^{(k)}=\frac{\mu_0\gamma_e\gamma_\mathrm{C}}{4\pi |\vec{r}^{(k)}|^3},
\end{equation}
$\mu_0$ the magnetic permeability of vacuum,
$\vec{r}^{(k)}=(r^{(k)}\sin\theta^{(k)}\cos\phi^{(k)},r^{(k)}\sin\theta^{(k)}\sin\phi^{(k)},r^{(k)}\cos\theta^{(k)})$
the displacement vector toward the $k^\mathrm{th}$ nucleus, and $\vec{e}^{(k)}$ the
unit vector of $\vec{r}^{(k)}$.

Additionally, since the $\widehat{S}_z$ component is responsible for $T_2^\ast$, while the $\widehat{S}_x$ and $\widehat{S}_y$ components are for $T_1$, the experimentally measured
three-order of magnitude difference between $T_1$ and $T_2^\ast$ guarantees a well-approximated pure dephasing of electron spin dynamics, on the time scale under study
\cite{redman_nv_relaxation_prl_1991,neumann_nv_entanglement_science_2008,childress_nv_center_science_2006,liu_nv_center_fid_sr_2012,maze_nv_center_fid_njp_2012}.
Therefore, it is relevant for us to neglect the terms proportional to $\widehat{S}_x$ and $\widehat{S}_y$ in Eq.~(\ref{eq_int_hamiltonian}) and consider only the $\widehat{S}_z$ component
phenomenologically.
Meanwhile, assuming that an external magnetic field $\vec{B}=B_z\vec{e}_z$ is aligned with the $z$-axis, then the total Hamiltonian can be expressed as
\begin{equation}
\widehat{H}_\mathrm{T}=D\widehat{S}^2_z+\gamma_eB_z\widehat{S}_z+\sum_k\gamma_\mathrm{C}B_z\widehat{J}_z^{(k)}+\widehat{S}_z\sum_k\vec{A}_z^{(k)}\cdot\widehat{J}^{(k)}.
\label{eq_total_hamiltonian}
\end{equation}
And the three components of the hyperfine coefficients are explicitly written as
\begin{equation}
\left\{\begin{array}{l}
A_{zx}^{(k)}=\alpha^{(k)}\left(-3\cos\theta^{(k)}\sin\theta^{(k)}\cos\phi^{(k)}\right) \\
A_{zy}^{(k)}=\alpha^{(k)}\left(-3\cos\theta^{(k)}\sin\theta^{(k)}\sin\phi^{(k)}\right) \\
A_{zz}^{(k)}=\alpha^{(k)}\left(1-3\cos^2\theta^{(k)}\right) \\
\end{array}\right..
\end{equation}

Due to the external magnetic field $\vec{B}=B_z\vec{e}_z$ lifting the $m_S=\pm 1$ degeneracy, now we selectively focus on the single-spin transition $\ket{0}\leftrightarrow\ket{1}$, forming a logical
qubit. With this setup, the total Hamiltonian (\ref{eq_total_hamiltonian}) is block diagonal in the electron spin basis in the form of
\begin{equation}
\widehat{H}_\mathrm{T}=\sum_{m_S=0,1}\ket{m_S}\bra{m_S}\otimes\widehat{H}_{m_S},
\label{eq_total_hamiltonian_qubit_manifold}
\end{equation}
with $\widehat{H}_{m_S}=(m_S^2D+m_S\gamma_eB_z)+\sum_k\vec{\Omega}_{m_S}^{(k)}\cdot\widehat{J}^{(k)}$, $\vec{\Omega}_{0}^{(k)}=\vec{\Omega}_{0}=(0,0,\gamma_\mathrm{C}B_z)$, and
$\vec{\Omega}_{1}^{(k)}=(A_{zx}^{(k)},A_{zy}^{(k)},A_{zz}^{(k)}+\gamma_\mathrm{C}B_z)$. Consequently, the corresponding total unitary time evolution operator
\begin{equation}
\widehat{U}_\mathrm{T}(t)=\exp(-i\widehat{H}_\mathrm{T}t)=\sum_{m_S=0,1}\ket{m_S}\bra{m_S}\otimes\widehat{U}_{m_S}(t),
\end{equation}
is also block diagonal in the electron spin basis with conditional evolution operators $\widehat{U}_{m_S}(t)=\exp(-i\widehat{H}_{m_S}t)$.

\subsection{Pure dephasing dynamics of electron spin and nonclassicality}

We now focus on the pure dephasing caused by the $^{13}$C nuclear spin bath during the free-induction-decay (FID) process. The initial state of total system is assumed to be a direct product of
all subsystems
\begin{equation}
\rho_\mathrm{T}(0)=\rho_\mathrm{NV}(0)\otimes\prod_k\rho^{(k)},
\end{equation}
where $\rho^{(k)}=[\widehat{I}^{(k)}+\vec{p}^{(k)}\cdot\hat{\sigma}^{(k)}]/2$ is the initial state of the $k^\mathrm{th}$ nuclear spin with polarization $\vec{p}^{(k)}$, and $\widehat{I}^{(k)}$ and
$\hat{\sigma}^{(k)}$ are the identity and the Pauli operators, respectively, acting on the $k^\mathrm{th}$ nuclear spin Hilbert space. After being optically polarized to $\ket{0}$, the initial state of
the electron spin is typically set to a balanced superposition $(\ket{0}+\ket{1})/\sqrt{2}$ by a $\pi/2$ MW pulse in a conventional FID experiment. Then the electron-nucleus hyperfine
interaction is turned on and the total system evolves unitarily according to $\rho_\mathrm{T}(t)=\widehat{U}_\mathrm{T}(t)\rho_\mathrm{T}(0)\widehat{U}_\mathrm{T}^\dagger(t)$, while the electron spin
reduced density matrix $\rho_\mathrm{NV}(t)=\Tr_\mathrm{C}\rho_\mathrm{T}(t)$ is obtained by tracing over the $^{13}$C nuclear spin bath.

The electron spin pure dephasing dynamics is characterized by the dephasing factor
\begin{equation}
\phi(t)=\bra{0}\rho_\mathrm{NV}(t)\ket{1}=e^{i(D+\gamma_eB_z)t}\prod_k\Tr\left[\widehat{U}_1^{(k)\dagger}(t)\widehat{U}_0^{(k)}(t)\rho^{(k)}\right],
\label{eq_dephasing_factor_ope_form}
\end{equation}
where $\widehat{U}_0^{(k)}(t)=\exp[-i(\vec{\Omega}_0\cdot\hat{\sigma}^{(k)})t/2]$ and $\widehat{U}_1^{(k)}(t)=\exp[-i(\vec{\Omega}_1^{(k)}\cdot\hat{\sigma}^{(k)})t/2]$ gives rise to nuclear spin
precession about the axis $\vec{u}^{(k)}=\vec{\Omega}_{1}^{(k)}/|\vec{\Omega}_{1}^{(k)}|$ in the presence of the hyperfine field produced by the electron spin. Further details for the calculation
of Eq.~(\ref{eq_dephasing_factor_ope_form}) are shown in \ref{app_dephasing_dynamics}.

In view of Eq.~(\ref{eq_pure_dephasing}), to construct a simulating HE for the electron spin pure dephasing, each member Hamiltonian in the ensemble is of the form $\omega\hat{\sigma}_z/2$ and the
CHER $\wp(\omega)$ is determined by the inverse Fourier transform
\begin{equation}
\wp(\omega)=\frac{1}{2\pi}\int_{-\infty}^\infty \phi(t) e^{i\omega t}dt.
\label{eq_inv_f_transform}
\end{equation}
From Eq.~(\ref{eq_inv_f_transform}), it is clear that the leading factor $\exp[i(D+\gamma_eB_z)t]$ on the right hand side of Eq.~(\ref{eq_dephasing_factor_ope_form}) merely shifts $\wp(\omega)$
by a displacement $D+\gamma_eB_z$, doing nothing to the profile of $\wp(\omega)$ nor to the nonclassical signatures. Besides, we are interested in the effects caused by the nuclear spin bath while
the leading factor is given by the energy space between the electron $\ket{0}$ and $\ket{1}$ states. Consequently, for our purpose, we can neglect the leading factor of
Eq.~(\ref{eq_dephasing_factor_ope_form}) and explicitly write down the dephasing factor as
\begin{eqnarray}
\phi(t)=\prod_k\left[\left(\cos\frac{\Omega_0t}{2}-ip_z^{(k)}\sin\frac{\Omega_0t}{2}\right)\cos\frac{\Omega_1^{(k)}t}{2}
+u_z^{(k)}\left(\sin\frac{\Omega_0t}{2}+ip_z^{(k)}\cos\frac{\Omega_0t}{2}\right)\sin\frac{\Omega_1^{(k)}t}{2}\right. \nonumber\\
\left.+i\left(p_x^{(k)}u_x^{(k)}+p_y^{(k)}u_y^{(k)}\right)\cos\frac{\Omega_0t}{2}\sin\frac{\Omega_1^{(k)}t}{2}
+i\left(p_x^{(k)}u_y^{(k)}-p_y^{(k)}u_x^{(k)}\right)\sin\frac{\Omega_0t}{2}\sin\frac{\Omega_1^{(k)}t}{2}\right].
\label{eq_dephasing_factor_use_this_form}
\end{eqnarray}

Since the CHER $\wp_\lambda$ is faithful for pure dephasing\cite{hongbin_process_n_cla_nc_2019,hongbin_cher_sr_2021}, we can construct a quantitative measure of nonclassicality in accordance
with the uniqueness of CHER for pure dephasing. Additionally, it is manifest that the classical CHERs form a convex set, i.e., the statistical mixture of classical CHERs is again a classical CHER,
therefore an intuitive measure can be defined as the distance from a nonclassical $\wp_\lambda$ to the classical set $\mathcal{C}$ of the conventional probability distribution $p_\lambda$, which is
given by \cite{hongbin_process_n_cla_nc_2019}
\begin{equation}
\mathcal{N}\{\mathcal{E}_t\} = \inf_{p_\lambda\in\mathcal{C}}\int_\mathcal{G}\frac{1}{2}|\wp_\lambda-p_\lambda| d\lambda.
\label{eq_measure_nonclassicality}
\end{equation}

\subsection{Nuclear spin polarization and precession}

From Eq.~(\ref{eq_dephasing_factor_use_this_form}), it is manifest that the polarization $\vec{p}^{(k)}$ and the precession axis of the nuclear spin, described by $\vec{u}^{(k)}$, have significant
influences on the electron spin dephasing dynamics. Therefore, it is possible to manipulate the dynamical behavior of the electron spin showing the transition between classicality and nonclassicality by
engineering the nuclear spin bath.

\begin{figure}[!ht]
\centering
\includegraphics[width=\textwidth]{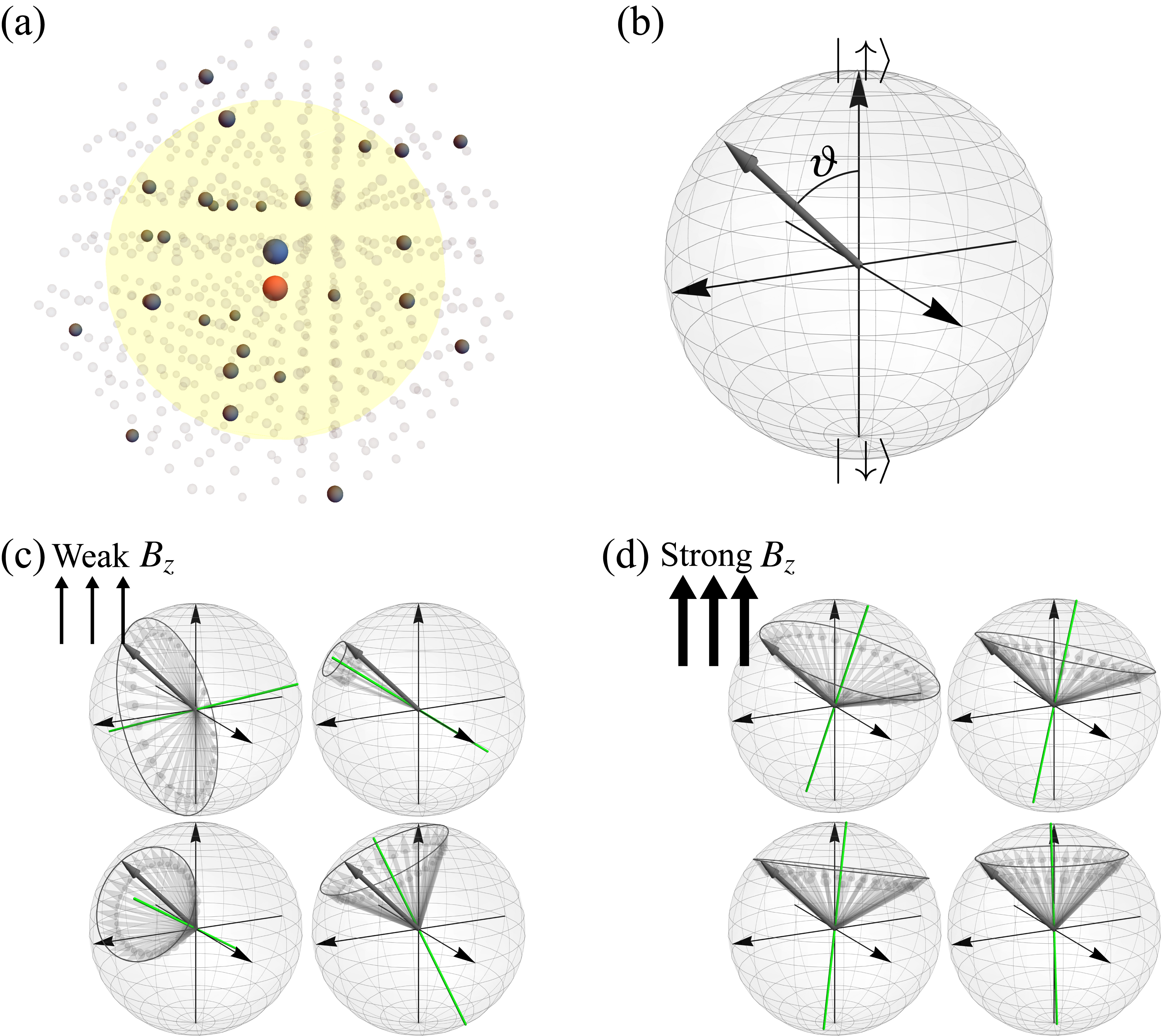}
\caption{Polarization and precession of the nuclear spin bath.
(a) DNP transfers electron spin polarization to the surrounding nuclear spins via the electron-nucleus hyperfine interaction. Only the nuclei within a polarization area (yellow spherical shadow)
can be efficiently polarized via direct polarization transfer, achieving hyperpolarization. Therefore, we assume a polarization area of radius 1 nm and only the nuclei within this area can be
identically polarized in a controllable manner. (b) We assume that the polarized nuclear spins within the polarization area are aligned in the $x$-$z$ plane with identical
$\vec{p}^{(k)}=(|\vec{p}|\sin\vartheta,0,|\vec{p}|\cos\vartheta)$. We will see that not only the magnitude $|\vec{p}|$, but also the orientation $\vartheta$ have significant influence on the electron
spin dynamics. (c) At weak fields, the axes of nuclear spin precessions (green axes) are randomly oriented due to the disordered $^{13}$C positions. (d) When the external field is
increased, most of the axes are gradually tilted and finally aligned regularly, approaching the $z$-axis since $\vec{\Omega}_{1}^{(k)}$'s are dominated by the external field.}
\label{fig_polarization_area}
\end{figure}

One of the mature approaches to engineer the bath is the dynamical nuclear polarization (DNP), which transfers the electron spin polarization to the surrounding nuclear spins via hyperfine interaction
and the resonance between them. Several approaches implementing DNP have been developed \cite{takahashi_nuc_spin_pola_prl_2008,
london_nuc_spin_pola_prl_2013,jacques_nuc_spin_pola_prl_2009,fischer_nuc_spin_pola_prl_2013,alvarez_nuc_spin_pola_nc_2015,king_nuc_spin_pola_nc_2015,scheuer_nuc_spin_pola_njp_2016,
chakraborty_nuc_spin_pola_njp_2017,scheuer_nuc_spin_pola_prb_2017,hovav_nuc_spin_pola_prl_2018,Henshaw18334_pnas_2019}.
Among these DNP approaches, the polarization mechanisms, as well as the resulting performances, differ from each other. Generically, it is not feasible to polarize the whole nuclear spin bath;
whereas, only a few number of nuclear spins within a polarization area, indicated by the yellow spherical shadow in figure~\ref{fig_polarization_area}(a), can be directly polarized and achieve
hyperpolarization. The rest of the nuclear spins outside the polarization area have a vanishingly low magnitude of polarization. Therefore, we assume that only the nuclei within 1 nm from the electron
spin possess identical and controllable polarization, i.e., finite $\vec{p}^{(k)}=(|\vec{p}|\sin\vartheta,0,|\vec{p}|\cos\vartheta)$ [figure~\ref{fig_polarization_area}(b)] for $\vec{r}^{(k)}\leq$ 1 nm;
otherwise $\vec{p}=0$. Moreover, not only the magnitude $|\vec{p}|$, but also the orientation $\vartheta$, are controllable.

The other one critical mechanism manipulating the FID classicality-nonclassicality transition is caused by the nuclear spin precession axes, which can be engineered by the external magnetic filed $B_z$.
This can be understood by observing that $\vec{\Omega}_{1}^{(k)}=\vec{A}_{z}^{(k)}+\gamma_\mathrm{C}B_z\vec{e}_z$, and $|\vec{A}_{z}^{(k)}|\propto|\vec{r}^{(k)}|^{-3}$. At weak fields, most of the
nuclear spin precession axes [green axes in figure~\ref{fig_polarization_area}(c)] are randomly oriented due to the randomly distributed $^{13}$C positions. Therefore, the electron spin will experience
a highly disordered hyperfine field caused by the randomly oriented nuclear spin precessions and, consequently, the nonclassical trait is smeared.

On the other hand, when the external field is increasing, the Zeeman splitting $\gamma_\mathrm{C}B_z\vec{e}_z$ gradually dominates most of $\vec{\Omega}_{1}^{(k)}$. Consequently, most of the axes will
tilt and finally align regularly, approaching the $z$-axis, i.e., $\vec{u}^{(k)}\approx\vec{e}_z$, as shown in figure~\ref{fig_polarization_area}(d), resulting in a more coherence hyperfine filed on the
electron spin.

\section{Numerical simulations}

In our numerical simulations, we have generated a configuration of nuclear spins consisting of 520 $^{13}$C nuclei randomly distributed over 47,231 lattice points, resulting in the natural abundance of
1.1$\%$. Additionally, to confirm the relevance of the dipole-dipole hyperfine interaction in the interaction Hamiltonian (\ref{eq_int_hamiltonian}), we have also verified that all $^{13}$C nuclei are
farther away than 5 {\AA} from the electron spin. After building an appropriately polarized nuclear spin bath, the magnetic field is set to be parallel to the $z$-axis at several different values.

\begin{figure}[!ht]
\centering
\includegraphics[width=\textwidth]{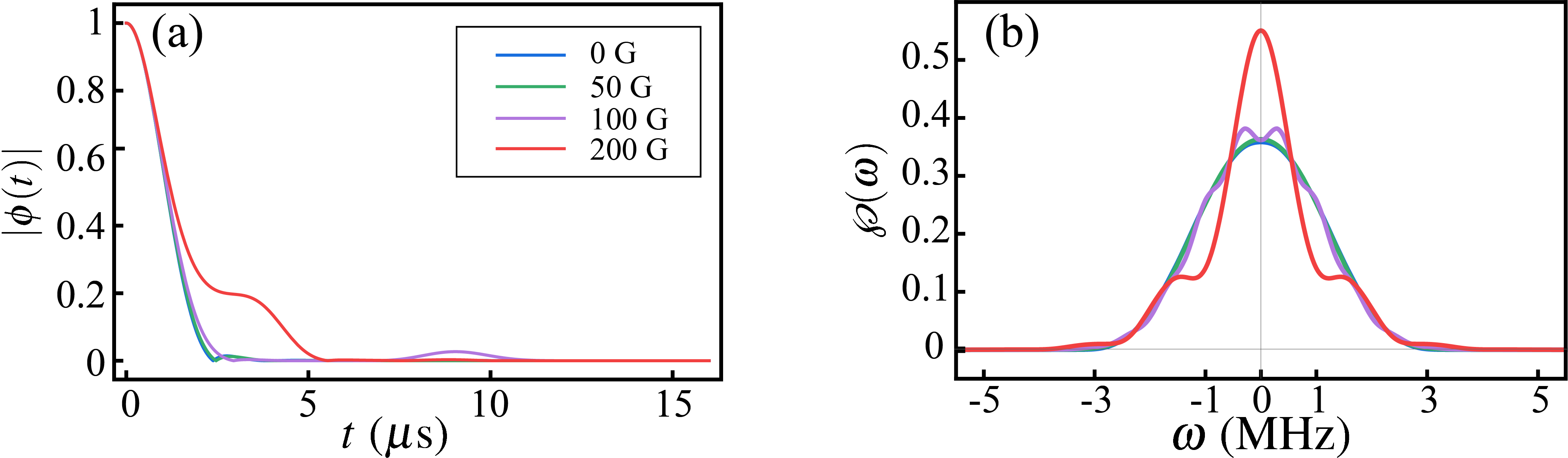}
\caption{Dynamics and CHER of the electron spin for an unpolarized nuclear spin bath.
(a) The dynamical behavior of dephasing factor for an unpolarized nuclear spin bath at various values of magnetic field. The sharp descent in the beginning becomes gentle at a large value of
magnetic field, indicating an enhanced $T_2^\ast$ time with increasing magnetic field. (b) The CHERs are symmetric since the nuclear spin bath is unpolarized. The positivity of the CHERs
indicates that the electron spin behaves classically in an unpolarized spin bath. Additionally, the crossover, from a smooth curve to wavy profile, when $B_z>100$~G is easily observed. This is a result
of the alignment of the tilted precession axes with increasing magnetic field.}
\label{fig_zero_pola}
\end{figure}

We first show the numerical results of Eq.~(\ref{eq_dephasing_factor_use_this_form}) in figure~\ref{fig_zero_pola} for an unpolarized nuclear spin bath, i.e., $|\vec{p}|=0$ for all nuclear spins, at
various values of the magnetic field. The dynamical behavior of the dephasing factor is shown in figure~\ref{fig_zero_pola}(a). We can observe that, when the magnetic field is large enough, the sharp
descent at the beginning becomes gentler, indicating an enhanced $T_2^\ast$ time. This is in good agreement with an experimental report \cite{maze_nv_center_fid_njp_2012}, wherein a similar explanation
in terms of a competition between $\vec{A}_{z}^{(k)}$ and $\gamma_\mathrm{C}B_z\vec{e}_z$ was proposed for the enhanced $T_2^\ast$. It is also intriguing to note that there exists a crossover when
$B_z>100$~G, which can be seen clearer from the profile of CHERs.

Figure~\ref{fig_zero_pola}(b) shows the corresponding CHERs $\wp(\omega)$, obtained from the inverse Fourier transform~(\ref{eq_inv_f_transform}). For the case of an unpolarized nuclear spin bath, the
profiles are symmetric and centered at $\omega=0$. The symmetry of the profiles can be understood from the viewpoint that an unpolarized spin is a mixture of two opposite polarizations of the same
magnitude, leading to two peaks at opposite positions as well as a symmetric CHER. In this case the CHERs are positive, indicating a classical-like dynamical behavior of electron spin.

\begin{figure}[!ht]
\centering
\includegraphics[width=\textwidth]{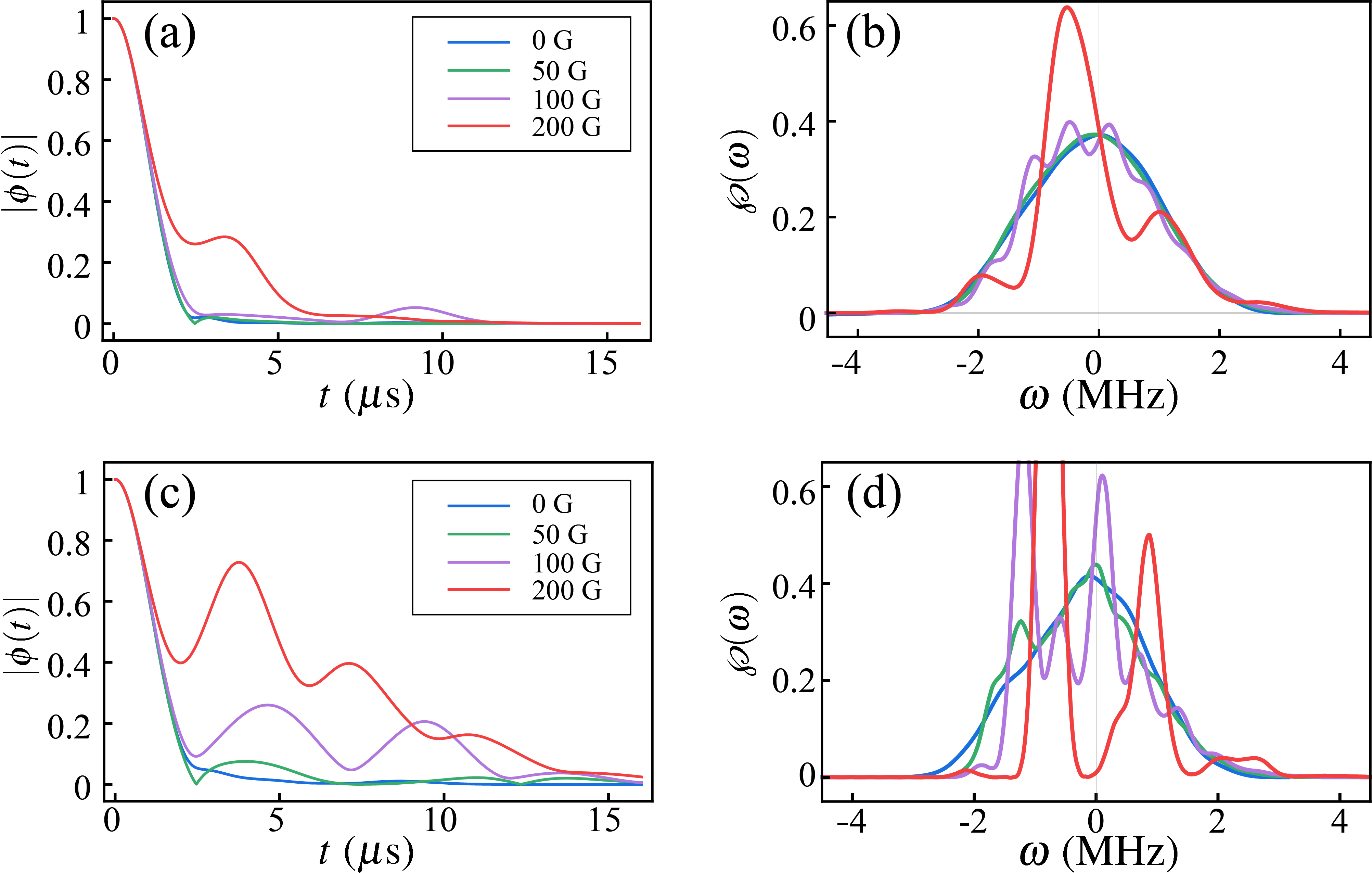}
\caption{Dynamics and CHER of the electron spin for a $z$-polarized nuclear spin bath.
The dynamical behavior (left panels) and the corresponding CHER (right panels) for the case of polarization toward the $z$-axis at magnitudes $|\vec{p}|=0.5$ [upper panels (a) and (b)] and
$1$ [lower panels (c) and (d)], respectively. From the dynamics shown in the left panels, we can observe that the sharp descent at the beginning is followed by a stronger oscillating tail
with increasing polarization magnitude $|\vec{p}|$, resulting in an enhanced $T_2^\ast$ time. This shows that the polarized nuclear spin toward the $z$-axis is capable of quenching the electron spin
decoherence. Meanwhile, the oscillating amplitude is increasing at strong fields, reflecting the almost regular alignment of tilted nuclear spin precession axes. From the CHER shown in the right panels,
the curves are pushed aside due to the polarized nuclear spin bath, leading to biased profiles. Notably, in the presence of both the hyper-$z$-polarization and the strong field, the CHERs are positive
without revealing a nonclassical trait.}
\label{fig_z_pola}
\end{figure}

Additionally, the aforementioned crossover is also clearer from the wavy profiles when $B_z>100$~G. The origin of this crossover can also be explained by the tilt of the precession axes illustrated in
the previous section. At weak fields, the profile is relatively smooth with less peaks, resulting from the randomly oriented precession axes. When the field is strong enough, the precession axes
gradually tilt regularly toward the $z$-axis. Particularly, when $B_z>100$~G, even the nuclear spins within the polarization area, which have dominant impact on the electron spin, gradually tilt as
well. Therefore, peaks emerge as the hyperfine fields caused by the tilted nuclear spins possess a consistent orientation, leading to wavy profiles.

We then proceed to investigate the impact of nuclear spin polarization on the electron spin dephasing dynamics. Figure~\ref{fig_z_pola} shows the results of polarization toward the $z$-axis, i.e.,
$\vartheta=0$, at magnitudes $|\vec{p}|=0.5$ (upper panels) and $1$ (lower panels), respectively. From the dynamical behavior of the dephasing factor shown in figures~\ref{fig_z_pola}(a) and (c), we
can observe that the oscillating tail following the sharp descent at the beginning becomes stronger with increasing polarization magnitude $|\vec{p}|$, resulting in an enhanced $T_2^\ast$ time as well.
This is also in line with experimental reports \cite{takahashi_nuc_spin_pola_prl_2008,london_nuc_spin_pola_prl_2013} that the polarized nuclear spin toward the $z$-axis is capable of quenching the
electron spin decoherence. Meanwhile, the oscillating amplitude is increasing at strong fields due to the alignment of the tilted nuclear spin precession axes, as schematically illustrated in
figures~\ref{fig_polarization_area}(c) and (d).

The impact of nuclear spin polarization is even prominent on the profile of CHER shown in figures~\ref{fig_z_pola}(b) and (d). In the presence of finite polarization, the mixture of two opposite
polarizations is no longer balanced, giving rise to a bias in the profile of the CHER. Moreover, the almost regularly aligned precession axes at strong fields render specific peaks even sharper. On the
other hand, it is worthwhile to note that, even if both the hyper-$z$-polarization and the strong field manipulate the profile of CHER significantly, those shown in figures~\ref{fig_z_pola} are still
positive without revealing a nonclassical trait.

\begin{figure}[!ht]
\centering
\includegraphics[width=\textwidth]{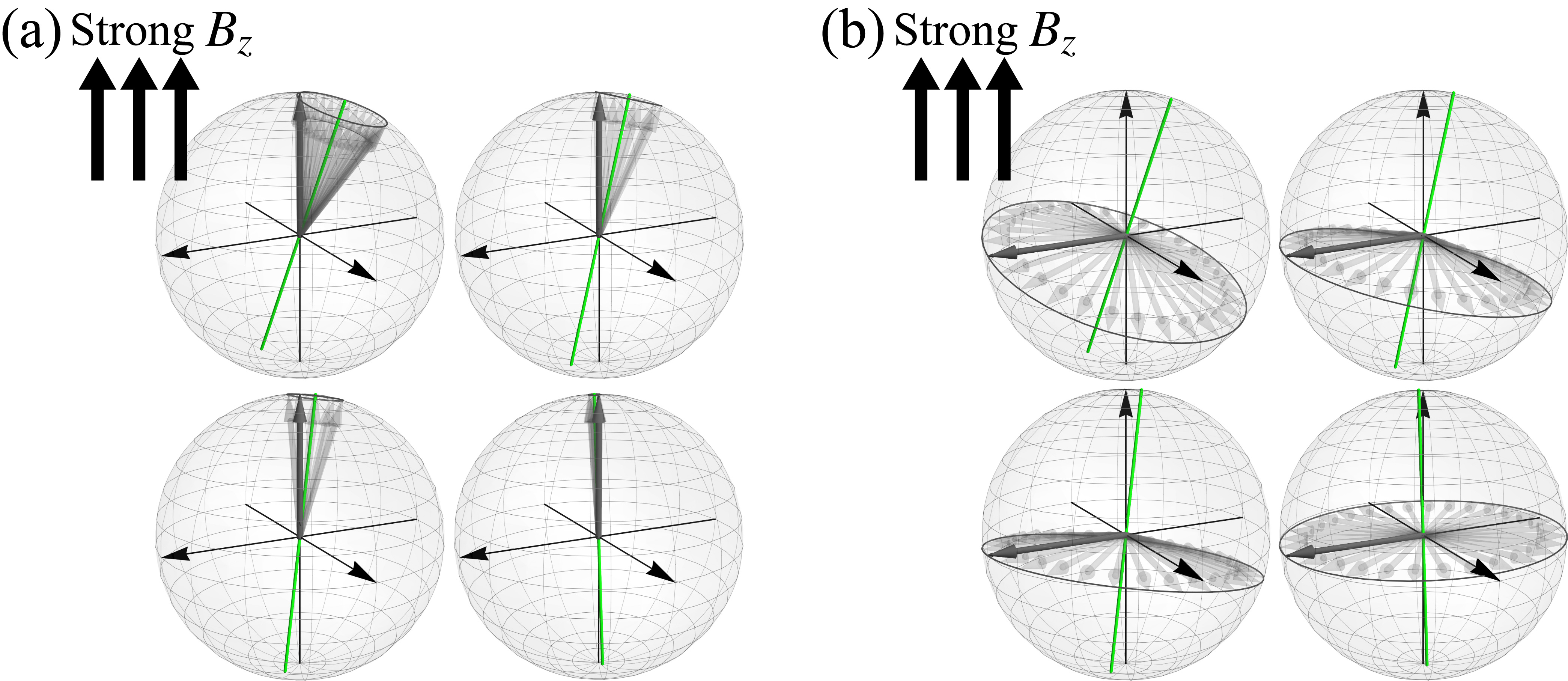}
\caption{Nuclear spin precession dynamics.
(a) For the case of polarization toward the $z$-axis, the almost regular alignment of the tilted nuclear spin precession axes at strong fields renders most of the angles between $\vec{u}^{(k)}$
and $\vec{p}$ very small. Therefore, most of the precession cones swept through by the nuclear spins are narrow, giving rise to a relatively static hyperfine field on the electron spin. (b) The
polarization toward the $x$-axis expands the precession cones, giving rise to a dynamic nuclear spin bath. The electron spin will behave nonclassically under such dynamic hyperfine field.}
\label{fig_alignment_prece_and_pola}
\end{figure}

We attribute this classicality to the extent of nuclear spin precession, which is a result of the alignment of the precession axes $\vec{u}^{(k)}$ and the orientation of the polarization $\vec{p}$. It
has been  illustrated in the previous section that the precession axes gradually tilt regularly toward the $z$-axis, i.e., $\vec{u}^{(k)}\approx\vec{e}_z$, with increasing magnetic field. Then, for the
case of polarization toward the $z$-axis, the alignment of the precession axes renders most of the angles between $\vec{u}^{(k)}$ and $\vec{p}$ very small, and, consequently, most of the precession
cones swept through by the nuclear spin are narrow [figure~\ref{fig_alignment_prece_and_pola}(a)]. Therefore, the electron spin will experience a relatively static hyperfine field caused by the
precessionless nuclear spin bath and behave classical-like. On the other hand, if the polarization is set toward the $x$-axis, the nuclear spin precession dynamics will be significantly different.
As shown in figure~\ref{fig_alignment_prece_and_pola}(b), the large angles between $\vec{u}^{(k)}$ and $\vec{p}$ will expand the precession cones, giving rise to a dynamic nuclear spin bath, as well as
a dynamic hyperfine field experienced by the electron spin. Consequently, the electron spin will reveal a prominent nonclassical trait.

\begin{figure}[!ht]
\centering
\includegraphics[width=\textwidth]{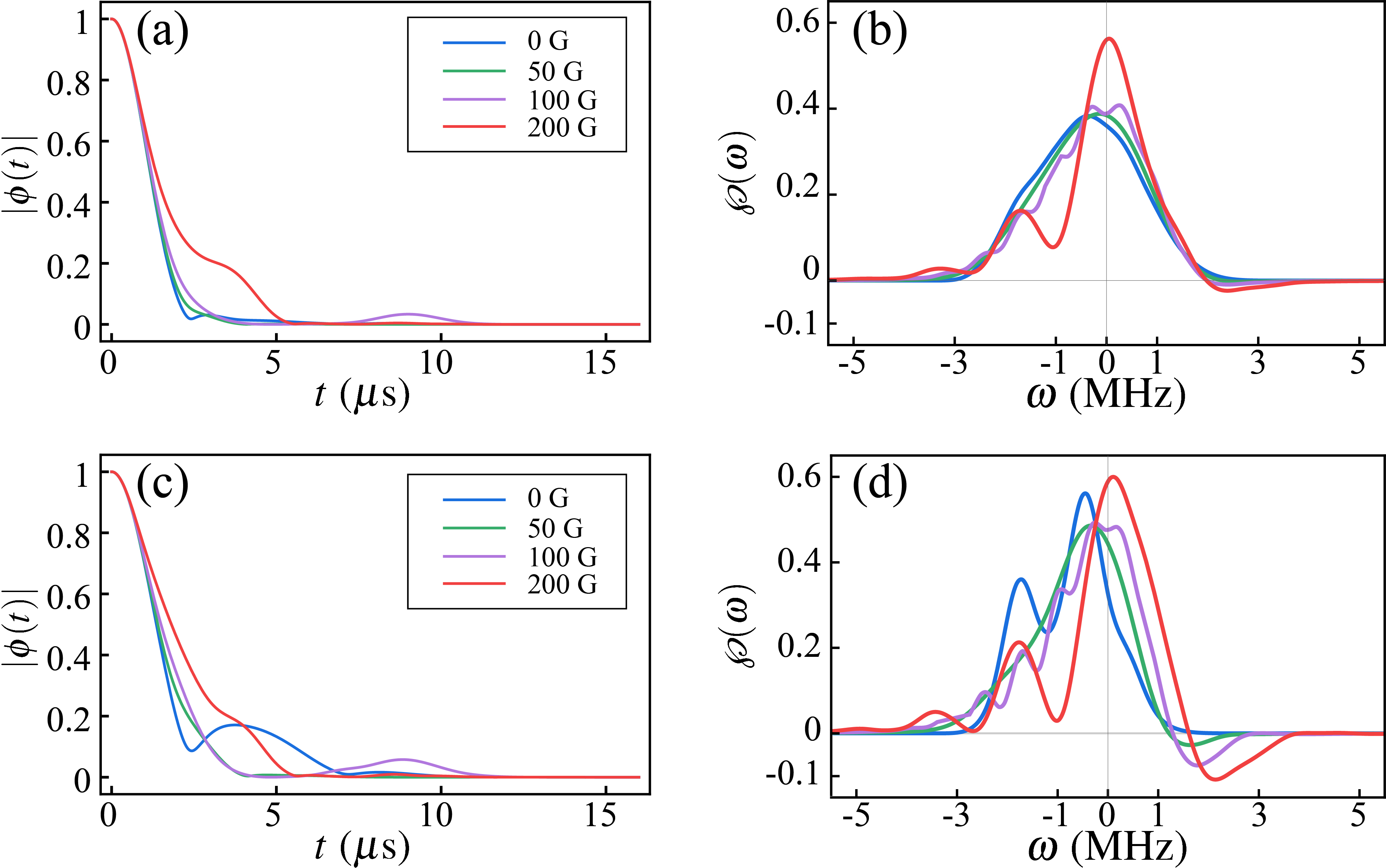}
\caption{Dynamics and CHER of the electron spin for an $x$-polarized nuclear spin bath.
The dynamical behavior (left panels) and the corresponding CHER (right panels) for the case of polarization toward the $x$-axis at magnitudes $|\vec{p}|=0.5$ [upper panels (a) and (b)] and
$1$ [lower panels (c) and (d)], respectively. The dynamical behavior shows a different response to the presence of $x$-polarization. It has negligibly small effects for prolonging the
$T_2^\ast$ time. Additionally, the dependence of the oscillating amplitude on the external magnetic field is nontrivial. On the contrary, the profiles of CHER show a qualitative difference from the case
of $z$-polarization. The asymmetry of the profiles raised by the $x$-polarization emerges in a manner of distortion. Additionally, the most exotic property is the emergence of negative values, which is
enhanced when increasing both $B_z$ and $|\vec{p}|$. This is the crucial indicator of the nonclassical trait of the electron spin dynamics caused by the nuclear spin precession dynamics.}
\label{fig_x_pola}
\end{figure}

To investigate the nonclassicality induced by the aforementioned nuclear spin precession dynamics, we assume that the nuclear spins are polarized toward the $x$-axis, i.e., $\vartheta=\pi/2$. The
numerical results are shown in figure~\ref{fig_x_pola} with magnitudes $|\vec{p}|=0.5$ (upper panels) and $1$ (lower panels), respectively. The dynamical behavior of the dephasing factor is shown in
figures~\ref{fig_x_pola}(a) and (c). In contrast to the case of $z$-polarization, the effect of prolonging the $T_2^\ast$ time by increasing the magnitude of the $x$-polarization is negligibly small.
Additionally, the dependence of the oscillating amplitude on the external magnetic field is also seemingly nontrivial. On the other hand, it is heuristic to note that, even if the dynamical behavior is
surely manipulated by the different types of polarization, generally speaking, the curves for different types of polarization share substantial similarities. In other words, the response of the curves
to the different types of polarization is not qualitatively sensitive.

On the contrary, the situation is very different for CHERs. As shown in figures~\ref{fig_x_pola}(b) and (d), rather than revealing a bias by pushing the curves aside caused by the $z$-polarization, the
asymmetry raised by the $x$-polarization emerges in a manner of distortion. This means that the profile of CHER reflects the difference in the two types of nuclear spin precession dynamics illustrated
in figure~\ref{fig_alignment_prece_and_pola}. Additionally, the most exotic property of the CHER raised by the $x$-polarization is the emergence of negative values, which is enhanced with both
increasing $B_z$ and $|\vec{p}|$. This, on the one hand, definitely certifies the nonclassicality of the electron spin pure dephasing dynamics in the presence of nuclear spin polarization toward the
$x$-axis; on the other hand, we also showcase the versatility of the CHER as a probe of nuclear spin bath dynamics.

\begin{figure}[!ht]
\centering
\includegraphics[width=\textwidth]{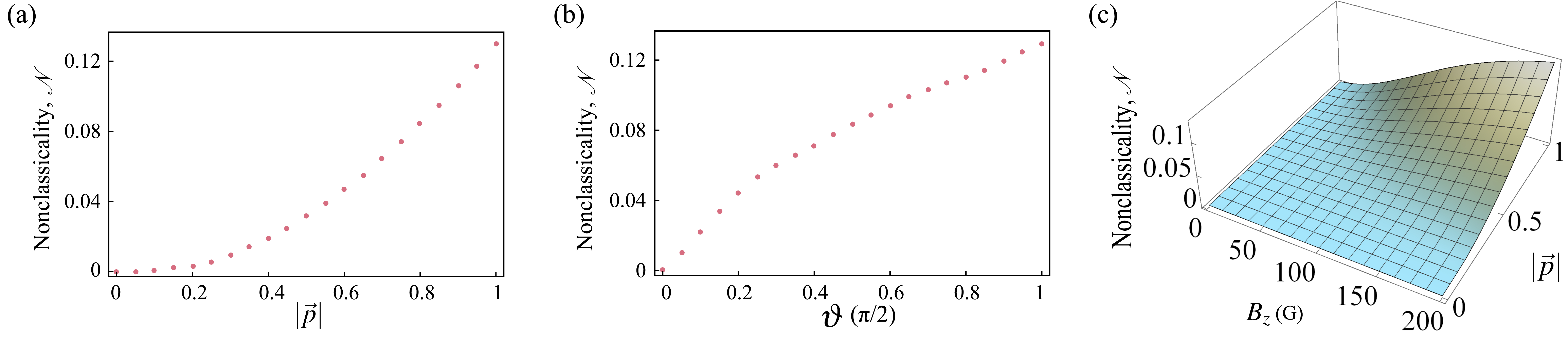}
\caption{Variation of nonclassicality with different parameters.
(a) The nonclassicality increases with magnitude $|\vec{p}|$ of the polarization. In this plot we have set an $x$-polarized nuclear spin bath ($\vartheta=\pi/2$) and $B_z=200$ G. (b) The nonclassicality
increases with the orientation $\vartheta$ of the polarization. Increasing $\vartheta$ indicates that the nuclear spin bath is rotated from the $z$-axis toward the $x$-axis. In this plot we have assumed
a hyperpolarization $|\vec{p}|=1$ and $B_z=200$ G. (c) This panel summarizes the overall response of the nonclassicality to the manipulation on the nuclear spin bath. The two experimentally
controllable parameters, $B_z$ and $|\vec{p}|$, denote two mechanisms of how we manipulate the nuclear spin precession dynamics, which in turn induces nonclassicality in the electron spin dephasing
dynamics. Therefore, the nonclassicality increases with $B_z$ and $|\vec{p}|$. In this plot we have set an $x$-polarized nuclear spin bath ($\vartheta=\pi/2$).}
\label{fig_numerical_nonclassicality}
\end{figure}

To quantitatively investigate the nonclassicality, we show the numerical results of nonclassicality $\mathcal{N}$ quantified by Eq.~(\ref{eq_measure_nonclassicality}) in
figure~\ref{fig_numerical_nonclassicality}. We first show the dependence on the magnitude $|\vec{p}|$ in figure~\ref{fig_numerical_nonclassicality}(a), where the polarization is set toward the $x$-axis
and $B_z=200$ G. The nonclassicality is increasing with $|\vec{p}|$, consistent with what we have seen from the CHERs in the presence of $x$-polarized nuclear spins [figures~\ref{fig_x_pola}(b) and
(d)]. Figure~\ref{fig_numerical_nonclassicality}(b) shows the dependence on the orientation $\vartheta$ with $|\vec{p}|=1$ and $B_z=200$ G. When $\vartheta=0$, the nuclear spins are $z$-polarized and
the electron spin dynamics reveals a classical-like behavior. When the nuclear spins are gradually rotated toward the $x$-axis, the nonclassicality increases due to the mechanism of nuclear spin
precession dynamics illustrated in figure~\ref{fig_alignment_prece_and_pola}(b). Finally, the overall response of the nonclassicality to the manipulation on the nuclear spin bath is shown in
figure~\ref{fig_numerical_nonclassicality}(c) with $\vartheta=\pi/2$. Both of the two experimentally controllable parameters $B_z$ and $|\vec{p}|$ manipulate the nuclear spin precession dynamics, which
in turn induces nonclassicality in the electron spin dephasing dynamics. Consequently, the nonclassicality increases with $B_z$ and $|\vec{p}|$.

\section{Experimental proposal}

Finally, in order to underpin the experimental viability of our numerical simulation, we also propose an experimental pulse sequence for carrying out the model. We stress that all the necessary
techniques included in this proposal are mature, up to an appropriate variation.

\begin{figure}[!ht]
\centering
\includegraphics[width=\textwidth]{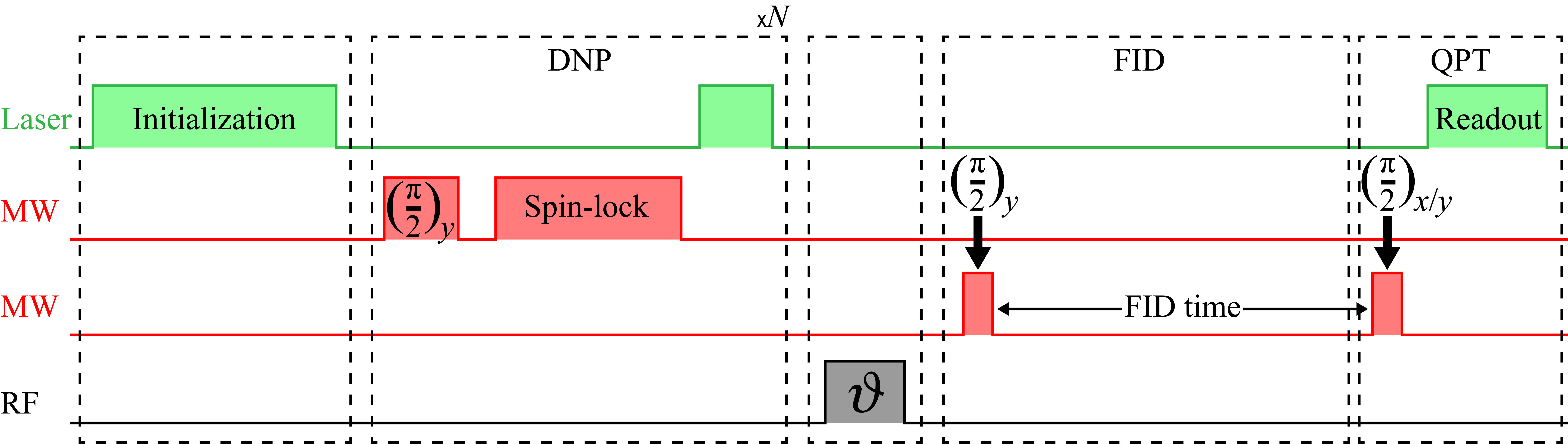}
\caption{The proposed experimental pulse sequence for carrying out the model. After the electron spin initialization to $\ket{0}$ by a 532-nm green laser,
a repeated DNP sequence followed by a RF pulse is used to build a hyperpolarized nuclear spin bath toward the orientation $\vartheta$.
Finally, a variant Ramsey sequence with two alternative final $(\pi/2)_{x/y}$ MW pulses can gather the signals of the imaginary and real parts of the dephasing
factor $\phi(t)$, respectively, fulfilling the requirement of QPT.}
\label{fig_exp_pulse_seq}
\end{figure}

Figure~\ref{fig_exp_pulse_seq} shows our proposal. The pulse sequence begins with an electron spin initialization to $\ket{0}$ by a 532-nm green laser. Then the DNP is applied to transfer the electron
spin polarization to the ambient nuclear spins. Several approaches implementing DNP have been developed \cite{takahashi_nuc_spin_pola_prl_2008,
london_nuc_spin_pola_prl_2013,jacques_nuc_spin_pola_prl_2009,fischer_nuc_spin_pola_prl_2013,alvarez_nuc_spin_pola_nc_2015,king_nuc_spin_pola_nc_2015,scheuer_nuc_spin_pola_njp_2016,
chakraborty_nuc_spin_pola_njp_2017,scheuer_nuc_spin_pola_prb_2017,hovav_nuc_spin_pola_prl_2018,Henshaw18334_pnas_2019}. A typical one, operating at a strong field with level anticrossing, begins with a
$(\pi/2)_y$ MW pulse rotating the electron spin about the $y$-axis to the $x$ direction. A following pulse locks the spin along the $x$ direction for a period, during which the electron spin
polarization will transfer to the ambient nuclear spins. The last step of the DNP sequence is an additional green laser pulse polarizing the electron spin again. The DNP sequence will be repeated $N$
times in order to build a hyperpolarized nuclear spin bath. After that, a radio-frequency (RF) pulse is applied to manipulate the orientation $\vartheta$. Finally, a Ramsey pulse sequence, operating at
desired fields, is used to activate the FID process of the electron spin.

Crucially, to implement the CHER theory, one should experimentally reconstruct the dynamical linear map $\mathcal{E}_t^{(\widehat{L})}$ in Eq.~(\ref{eq_ft_on_group}), which requires a full QPT
experiment to gather necessary information on the qubit dynamics. However, the conventional Ramsey sequence is clearly insufficient for QPT. Due to the pure dephasing dynamical behavior of the electron
spin, we propose a variant Ramsey sequence with two alternative final $(\pi/2)_{x/y}$ MW pulses before the optical readout. These two readout signals correspond to the imaginary and real parts of the
dephasing factor $\phi(t)$, respectively, fulfilling the requirement of QPT.

\section{Conclusions}\label{sec_conclusion}

In conclusion, we have analyzed the pure dephasing dynamics and the corresponding CHER with an authentic quantum system of an NV$^-$ center. By engineering the nuclear spin precession dynamics, on the
one hand, we can manipulate the dynamical behavior of the electron spin showing the transition between classicality and nonclassicality during the FID process; on the other hand, we have also
investigated the process nonclassicality from a new viewpoint of quantum-dynamical mechanism, rather than the original quantum-information-theoretic perspective. This reveals not only how the nuclear
spin precession dynamics gives rise to the nonclassical trait in the electron spin FID process, but also the role played by the environmental dynamics in the origin of dynamical process nonclassicality.

Following the logic of the violation of Bell's inequality or the negativity in the phase space representation of a bosonic field, the nonclassicality characterized by the CHER is based on
the failure of a classical strategy formulated in terms of HEs, which is shown to be closely related to the nonclassical correlations between the system and its environment. By further recasting the
ensemble-averaged dynamics under a HE into a Fourier transform using the formalism of group theory, the role played by the CHER as a characteristic representation of a dynamical process over the
frequency domain becomes manifest. Then we can quantitatively define the dynamical process nonclassicality in view of the negativity in the CHER.

We have applied the CHER theory to the FID process of the electron spin associated to an NV$^-$ center in the diamond lattice and discovered how the nonclassicality is induced by the nuclear spin
precession dynamics. There are two experimentally mature approaches engineering the nuclear spin precession dynamics, i.e., the external magnetic field and the DNP. The former tends to rotate the
precession axes via a competition with the randomly oriented hyperfine interaction. At strong fields, most of the precession axes are regularly aligned along the $z$-axis, resulting in a more coherent hyperfine filed experienced by the electron spin. While the latter transfers electron spin polarization to the surrounding nuclear spins within a polarization area and establishes hyperpolarization.

Here we have assumed a polarization area of radius 1 nm. For the case of polarization toward the $z$-axis at strong fields, most of the precession cones swept through by the nuclear spins are narrow,
giving rise to a relatively static hyperfine field on the electron spin. If the nuclear spin polarizations are rotated toward the $x$-axis, the precession cones are expanded, giving rise to a dynamic
nuclear spin bath. We found that the electron spin will behave nonclassically under a dynamic hyperfine field caused by the expanded precession cones.

This can be seen from the numerical simulations. The increasing magnetic field and $z$-polarization both can be used to enhance the $T_2^\ast$ time, in good agreement with experimental reports.
While the CHER will show a crossover from a smooth curve to wavy profile with increasing field and an asymmetry with larger polarization magnitude. Additionally, even in the presence of both the
hyper-$z$-polarization and the strong fields the CHER is positive, indicating a classical-like electron spin FID process. On the other hand, in the case of $x$-polarized nuclear spin bath, the dynamic
hyperfine field caused by the nuclear spin precession gives rise to prominent negativity in the corresponding CHER. However, the $x$-polarization is not capable of enhancing the $T_2^\ast$ time
significantly. Consequently, we conclude that the nonclassicality will be stronger with increasing magnetic field and $x$-polarization. Finally, we also present an experimental pulse sequence for
carrying out the model. Our proposal combines several mature techniques, including optical initialization and readout of electron spin, DNP, FID, and QPT.


\section*{Data availability statement}

The data that support the findings of this study are available upon reasonable request from the corresponding authors.

\section*{Acknowledgments}

This work is supported by the Ministry of Science and Technology, Taiwan, Grants No. MOST 108-2112-M-006-020-MY2, MOST 109-2112-M-006-012, MOST 110-2112-M-006-012, and MOST 111-2112-M-006-015-MY3,
and partially by Higher Education Sprout Project, Ministry of Education to the Headquarters of University Advancement at NCKU.
F.N. is supported in part by
Nippon Telegraph and Telephone Corporation (NTT) Research,
the Japan Science and Technology Agency (JST) [via the Quantum Leap Flagship Program (Q-LEAP), and the Moonshot R\&D Grant Number JPMJMS2061],
the Japan Society for the Promotion of Science (JSPS) [via the Grants-in-Aid for Scientific Research (KAKENHI) Grant No. JP20H00134],
the Army Research Office (ARO) (Grant No. W911NF-18-1-0358),
the Asian Office of Aerospace Research and Development (AOARD) (via Grant No. FA2386-20-1-4069), and
the Foundational Questions Institute Fund (FQXi) via Grant No. FQXi-IAF19-06.

\appendix \renewcommand{\thesection}{Appendix \Alph{section}}

\section{Derivation of the electron spin dephasing factor}\label{app_dephasing_dynamics}

Here we show how to obtain the expression for the dephasing factor~(\ref{eq_dephasing_factor_use_this_form}) from Eq.~(\ref{eq_dephasing_factor_ope_form}).
We consider the qubit manifold defined by the $\ket{0}\leftrightarrow\ket{1}$ transition. Then the block diagonal total Hamiltonian~(\ref{eq_total_hamiltonian_qubit_manifold}) leads to a block diagonal
unitary time evolution operator $\widehat{U}_\mathrm{T}(t)=\exp(-i\widehat{H}_\mathrm{T}t)=\ket{0}\bra{0}\otimes\widehat{U}_{0}+\ket{1}\bra{1}\otimes\widehat{U}_{1}$, where
\begin{equation}\left\{
\begin{array}{l}
\widehat{U}_1=\exp[-i(D+\gamma_eB_z)t]\prod_k\exp[-i(\vec{\Omega}_{1}^{(k)}\cdot\hat{\sigma}^{(k)})t/2] \\
\widehat{U}_0=\prod_k\exp[-i(\vec{\Omega}_0\cdot\hat{\sigma}^{(k)})t/2]
\end{array}\right.,
\end{equation}
where $\vec{\Omega}_{1}^{(k)}=(A_{zx}^{(k)},A_{zy}^{(k)},A_{zz}^{(k)}+\gamma_\mathrm{C}B_z)$, and $\vec{\Omega}_0=(0,0,\gamma_\mathrm{C}B_z)$. Then the electron spin reduced density matrix
\begin{equation}
\rho_\mathrm{NV}(t)=\Tr_\mathrm{C}[\widehat{U}_\mathrm{T}(t)\rho_\mathrm{T}(0)\widehat{U}_\mathrm{T}^\dagger(t)]
\end{equation}
is obtained by tracing over the $^{13}$C nuclear spin bath from the total density matrix.

Neglecting the internuclear initial correlations by considering the initial state $\rho_\mathrm{T}(0)=\rho_\mathrm{NV}(0)\otimes\prod_k\rho^{(k)}$ with the nuclear spin initial state
$\rho^{(k)}=[\widehat{I}^{(k)}+\vec{p}^{(k)}\cdot\hat{\sigma}^{(k)}]/2$, then the dephasing factor
\begin{equation}
\phi(t)=\bra{0}\rho_\mathrm{NV}(t)\ket{1}=\exp[i(D+\gamma_eB_z)t]\prod_k\Tr\left[\widehat{U}_1^{(k)\dagger}(t)\widehat{U}_0^{(k)}(t)\rho^{(k)}\right]
\end{equation}
is given by a product of the effect of each single nuclear spin. To calculate $\phi(t)$, recall that
\begin{equation}\left\{
\begin{array}{l}
\widehat{U}_1^{(k)}(t)=\exp[-i(\vec{\Omega}_1^{(k)}\cdot\hat{\sigma}^{(k)})t/2]=\cos(\Omega_1^{(k)}t/2)\widehat{I}^{(k)}-i\sin(\Omega_1^{(k)}t/2)(\vec{u}^{(k)}\cdot\hat{\sigma}^{(k)}) \\
\widehat{U}_0^{(k)}(t)=\exp[-i(\vec{\Omega}_0\cdot\hat{\sigma}^{(k)})t/2]=\cos(\Omega_0t/2)\widehat{I}^{(k)}-i\sin(\Omega_0t/2)(\vec{e}_z\cdot\hat{\sigma}^{(k)})
\end{array}\right..
\end{equation}
Additionally, with the help of the prescription $(\vec{u}\cdot\hat{\sigma})(\vec{v}\cdot\hat{\sigma})=(\vec{u}\cdot\vec{v})\widehat{I}+i(\vec{u}\times\vec{v})\cdot\hat{\sigma}$, we have
\begin{eqnarray}
\widehat{U}_1^{(k)\dagger}(t)\widehat{U}_0^{(k)}(t)&=&\left(\cos\frac{\Omega_0t}{2}\cos\frac{\Omega_1^{(k)}t}{2}+\sin\frac{\Omega_0t}{2}\sin\frac{\Omega_1^{(k)}t}{2}\vec{u}^{(k)}\cdot\vec{e}_z\right)\widehat{I}^{(k)}
+i\sin\frac{\Omega_0t}{2}\sin\frac{\Omega_1^{(k)}t}{2}(\vec{u}^{(k)}\times\vec{e}_z)\cdot\hat{\sigma}^{(k)} \nonumber\\
&&+i\cos\frac{\Omega_0t}{2}\sin\frac{\Omega_1^{(k)}t}{2}(\vec{u}^{(k)}\cdot\hat{\sigma}^{(k)})-i\sin\frac{\Omega_0t}{2}\cos\frac{\Omega_1^{(k)}t}{2}(\vec{e}_z\cdot\hat{\sigma}^{(k)}).
\end{eqnarray}
Due to the orthogonality of the identity and the Pauli operators $\Tr\hat{\sigma}_j\hat{\sigma}_k=2\delta_{jk}$, the trace taken over the Hilbert space of the $k^\mathrm{th}$ nuclear spin is easy to
perform. Then we obtain the desired result:
\begin{eqnarray}
\phi(t)=e^{i(D+\gamma_eB_z)t}\prod_k\left[\left(\cos\frac{\Omega_0t}{2}-ip_z^{(k)}\sin\frac{\Omega_0t}{2}\right)\cos\frac{\Omega_1^{(k)}t}{2}
+u_z^{(k)}\left(\sin\frac{\Omega_0t}{2}+ip_z^{(k)}\cos\frac{\Omega_0t}{2}\right)\sin\frac{\Omega_1^{(k)}t}{2}\right. \nonumber\\
\left.+i\left(p_x^{(k)}u_x^{(k)}+p_y^{(k)}u_y^{(k)}\right)\cos\frac{\Omega_0t}{2}\sin\frac{\Omega_1^{(k)}t}{2}
+i\left(p_x^{(k)}u_y^{(k)}-p_y^{(k)}u_x^{(k)}\right)\sin\frac{\Omega_0t}{2}\sin\frac{\Omega_1^{(k)}t}{2}\right].
\end{eqnarray}
Finally, neglecting the leading factor leads to Eq.~(\ref{eq_dephasing_factor_use_this_form}).






\end{document}